\documentclass[aps,pre,onecolumn,12pt,a4paper,showpacs,dvips,floatfix]{revtex4}
\usepackage{graphicx}

\begin{document}

\title{Spectral statistics of the quenched normal modes of a network-forming molecular liquid}

\author{\bf Gurpreet S. Matharoo, M. Shajahan G. Razul and Peter H. Poole}

\affiliation{Department of Physics, St. Francis Xavier University, Antigonish,
Nova Scotia B2G 2W5, Canada}

\date{\today\,  -- Manuscript A08.09.0208 -- Resubmitted Version}

\begin{abstract}
We evaluate the density of states of the quenched normal modes of ST2 water, and their statistical fluctuations, for a range of densities spanning three regimes of behavior of a hydrogen bonded liquid:  a lower-density regime of random tetrahedral network formation; in the vicinity of a liquid-liquid critical point; and in a higher-density regime of fragile glass-forming behavior.  For all cases we find that the fluctuations around the mean spectral densities obey the predictions of the Gaussian orthogonal ensemble of random matrix theory.  We also measure the participation ratio of the normal modes across the entire frequency range, and find behavior consistent with the majority of modes being of an extended nature, rather than localized.
\end{abstract}


\maketitle

\section{Introduction}

The nature of the potential energy surface (PES) of liquids and glasses has attracted significant attention over the past decade.   Numerous computer simulation studies have successfully demonstrated that both the thermodynamic and transport behavior of glass-forming liquids can be better understood through a consideration of the PES, in particular by examining its local minima, or inherent structures (IS), and the nature of the energy barriers that separate them~\cite{p01,p02,p03,p04,p05,p06,p07,p08,p09,p10,p11,p12,p13,p14,p15,p16,p17,p18,p19,p20,
p21a,p21b,p22,schir,gsm}.

A fundamental property describing the topology of the PES at a particular point in configuration space is the matrix of second derivatives (or Hessian matrix) of the potential energy with respect to the degrees of freedom of the system.  The eigenvalues and eigenvectors of the Hessian matrix characterize the normal modes of the configuration under consideration.   If the configuration is a snapshot of an equilibrium liquid, the modes are called instantaneous normal modes (INM), and the spectrum of eigenvalues will contain both positive and negative values.  If the configuration is an IS, the modes are called quenched normal modes (QNM), in which case all the eigenvalues are positive.  Given an ensemble of IS's evaluated from equilibrium liquid configurations, the equilibrium probability density for QNM's of a given frequency can be found, and corresponds to the vibrational density of states of the system in the harmonic approximation.

While the shape of vibrational spectra of disordered systems in general can vary widely from one system to another, commonalities have been discovered using concepts from random matrix theory~\cite{p10,p11,p12,p13}.  For example, random matrix theory has been used extensively for the analysis of statistical fluctuations of the spectra of complex quantum systems such as complex nuclei,  
quantum chaotic systems and disordered mesoscopic systems~\cite{p28,p29,p30}.  Study of the spectra of these systems has revealed that the fluctuations around mean spectral densities can always be assigned to one of three universality classes, even though mean spectral densities themselves may be highly system dependent.  

Notably, in the context of glass-forming systems, recent simulation studies conducted on the vibrational spectra of several liquid and amorphous solid systems have shown that the fluctuation properties of the vibrational spectra are described by the Gaussian orthogonal ensemble (GOE) of random matrices~\cite{p14,p13,p10,p11,p12,gsm}.  These results have been obtained using spectra derived from both INM and QNM properties, although the clearest indications seem to have come from using QNM data.  All the glass-forming systems tested to date have been found to obey GOE statistics~\cite{FN1}.  These systems make use of various spherically-symmetric atomic interaction potentials and include both bulk liquids and glasses as well as clusters, in both two and three dimensions.  The results suggest that the statistics of the fluctuations of vibrational spectra in all liquids and glasses may obey a common pattern, represented by the GOE universality class.  

The identification of universality in the fluctuations of liquid vibrational spectra can assist in checking analytical approaches (see e.g. Refs.~\cite{stratt2,check}) that attempt to evaluate densities of state directly from a system Hamiltonian, a fundamental challenge of liquid-state physics.  The nature of the universality also informs our understanding of the dynamical processes in liquids and glasses.  If the universality is that of the GOE of random matrix theory, then the implication is that the normal modes of the system are extended (rather than localized)~\cite{schir}, thus elucidating the nature of the system's motion on the PES.

In this work, we address the question of whether this proposed universality extends beyond atomic systems to molecular, network-forming liquids.  This is an important class of glass-forming systems, as it includes water, silica and many other materials that form the basis of amorphous solids in a range of contexts, from biology to earth sciences.  Here, we will employ computer simulations of a model water potential to represent the class of network-forming liquids in which tetrahedral coordination dominates at the local level.  As described below, this will allow us to extend the search for GOE behavior to a range of conditions not accessible using simpler atomic potentials.

\section{Methods}

To model water, we use the ST2 pair potential formulated by Stillinger and Rahman~\cite{p23}. Over the years, there have been numerous studies that have used this model to elucidate the complex phenomena found in water, and other liquids having a local tetrahedral molecular structure; see e.g. Refs.~\cite{p25,p24,BGO,p27,limei,p26}, and references therein.  While other simulation models exist that more accurately reproduce water properties, ST2 has the advantage of displaying a diversity of extreme liquid behavior in one system.  With regard to the PES of water, the initial study on the IS in water was conducted by Stillinger and Weber~\cite{p04}. Thereafter there have been numerous studies of liquid water dynamics and its relationship to the underlying PES using various model potentials~\cite{p17,p18,p19,p20,p21a,p21b,p22}.

We study a periodic ST2 system with a constant number of molecules $N = 1728$ (unless otherwise noted), constant volume $V$, and with temperature $T$ controlled by a Berendsen thermostat~\cite{beren}. We conduct conventional molecular dynamics simulations~\cite{AT} using the ``leap-frog" algorithm with a $1$~fs time step, and where the long range nature of the ST2 interaction potential is approximated using the reaction field method, with direct intermolecular interations cut off at 0.78~nm.  Our molecular dynamics code uses the SHAKE algorithm to constrain the motion of the ST2 molecule to that of a rigid body~\cite{shake}.

We study three densities, $\rho = 0.83$, $0.93$ and $1.09$~g/cm$^3$ at $T = 260$~K.  These three densities are chosen (following the reasoning of Becker, et al.~\cite{p26}) to allow us to explore a wide range of liquid behavior in ST2 water.  A tetrahedral network of hydrogen bonds is predominant throughout this density range.  At the same time, changes in the hydrogen bond network with density are correlated to qualitative changes in thermodynamic and transport properties.  At the lowest density we study, $\rho = 0.83$~g/cm$^3$, the behavior of ST2 water is characterized by the emergence of an increasingly defect-free random tetrahedral network as $T$ decreases.  This is accompanied by the detection of both a density maximum and a density minimum~\cite{p27}, and by a fragile-to-strong dynamical crossover~\cite{limei}, by following the behavior along this isochore.  At the intermediate density of $\rho = 0.93$~g/cm$^3$, we enter the regime at which a liquid-liquid phase separation (between a low density liquid and a high density liquid phase) occurs below approximately $T=245$~K~\cite{p25}.  At $T=260$~K, the system is under the influence of critical fluctuations associated with the critical point of this phase transition.  Finally, at the highest density we study, $\rho=1.09$~g/cm$^3$, the tetrahedral structure of the first coordination shell is increasingly disrupted by the penetration of a non-hydrogen-bonded fifth neighbor molecule, inducing dynamical behavior more like that of a fragile glass-former~\cite{limei}. 

In the following, we analyze the vibrational density of states of the QNM's of the system, and the statistics of their fluctuations, at each of these densities. The three distinct regimes of liquid-state behavior represented by these densities (respectively: random tetrahedral network, critical fluctuations, and fragile dynamics) provide a range of extreme thermodynamic and dynamical conditions in a network-forming molecular liquid for testing the robustness of the universal behavior of vibrational spectra that have been reported for simpler atomic systems. Our aim is to test if any of these complexities of a tetrahedral network is sufficient to induce an exception to this universality.  

To simplify the comparison of our results at different densities, our data is generated at $T=260$~K.  This is low enough to reveal the characteristic behavior represented by the three densities chosen.  We have also checked that our main conclusions do not change if the analysis at each density is carried out at $300$~K.  It is important to note that we do not expect our results to depend on whether or not the liquid is supercooled at these $T$.  We choose to study this low $T$ regime simply to ensure that the network-forming character of the liquid is prominent.

We also analyze the normal modes to determine if they are extended (i.e. involve an extensive number of particles of the system), or localized to a finite cluster of molecules.  Previous work has connected extended modes with those obeying GOE statistics, while localized ones are thought to be consistent with Poissonian statistics~\cite{schir}.  Previous studies of network-forming liquids, including silica and water,  have reported that a majority of modes are extended in nature~\cite{p16,p19}.  We study this issue here by evaluating the participation ratios~\cite{p09,p10,p11,p12,p13,p14,p15,p16,p19,p20,gsm} of the QNM's, to test if their behavior in this regard is consistent with the behavior found for the statistics of the spectral fluctuations.

\section{Density of states}

\begin{figure}
\centerline{\includegraphics[width=3.0in]{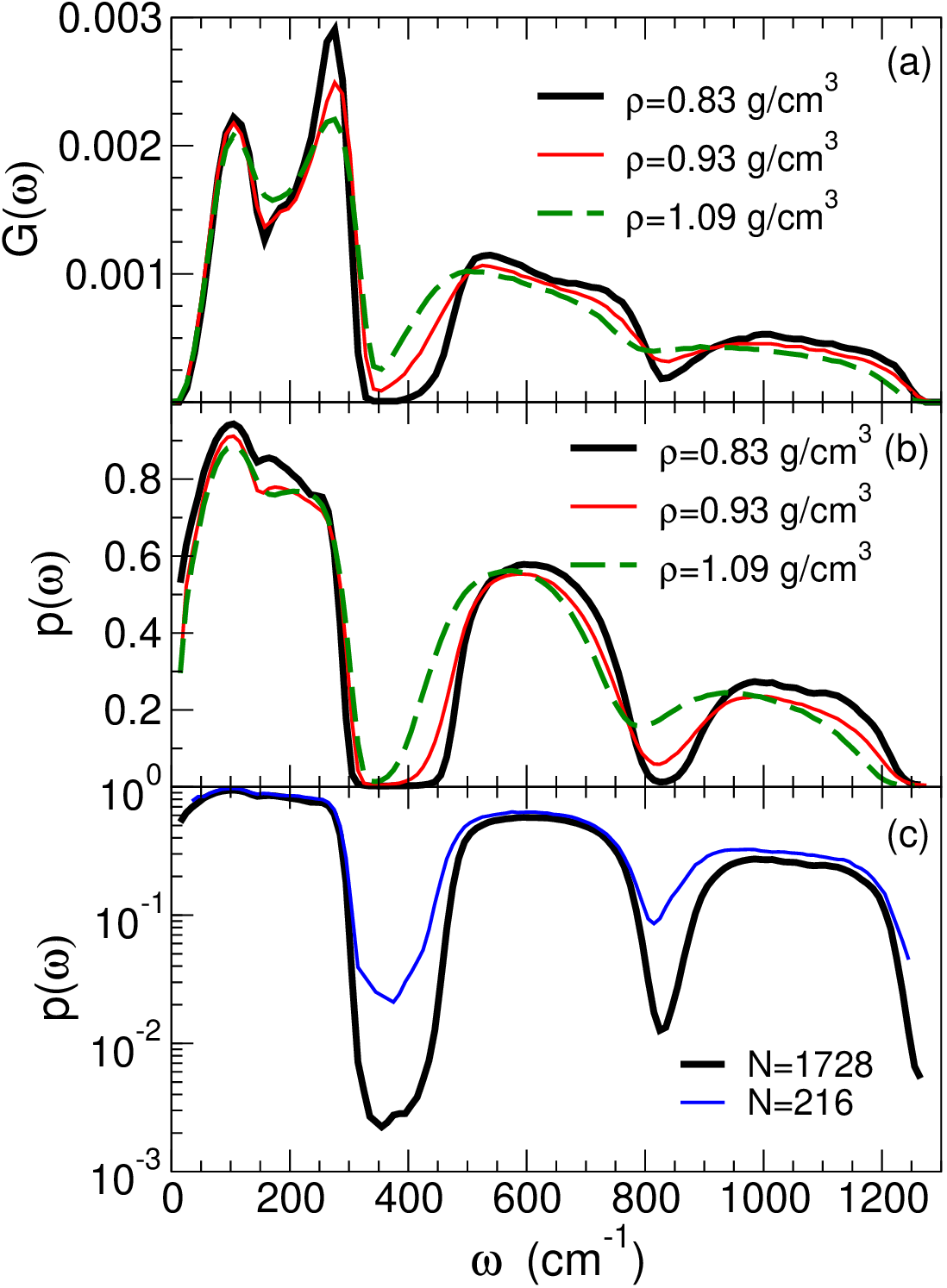}}
\caption{(a) DOS versus $\omega$ of quenched normal modes plotted for various
densities at $T = 260$~K and for $N=1728$. The plot for each density is an average over 100 configurations, and is normalized to unit area. (b) Average participation ratios versus $\omega$ of the quenched normal modes for various densities at $T = 260$~K and $N=1728$. The plot for each density is averaged over 100 configurations. (c) Average participation ratios versus $\omega$, plotted on a semi-log scale, of the quenched normal modes for $\rho = 0.83$~g/cm$^3$ and $T=260$~K obtained from simulations of two different system sizes, $N=216$ and $1728$ molecules.}
\label{fig1}
\end{figure} 

We evaluate the density of states (DOS) from liquid configurations harvested from our simulations.  For each density, the system is allowed to equilibrate with the temperature controlled by a Berendsen thermostat.  The equilibration runs at each $T$ are carried out for the time required for the mean squared displacement (MSD) to reach 1~nm$^2$, or for 100 ps, whichever is larger. Note that a MSD of 1~nm$^2$ corresponds to an average displacement of about 3 times the distance between nearest neighbour molecules.  This equilibration criterion is the same as that used in Ref.~\cite{p27} and produces stable and reproducible time series of thermodynamic quantities.  After equilibrium is attained, the runs are continued, again with $T$ controlled using a Berendsen thermostat.  From these production runs we generate 100 distinct configurations, separated in time by 10~ps.  We then quench each configuration to its corresponding IS using the conjugate gradient method. 

For each IS, we evaluate the Hessian matrix with Bloch wave vector $k =0$.  The definition of the Hessian matrix used here is exactly the same as that given in Appendix~B of Ref.~\cite{p21a}.  We then diagonalize the Hessian matrix to obtain both the eigenvalues~$\left(\lambda\right)$ as well as the eigenvectors.   The eigenvectors correspond to the normal modes of the IS configuration and the eigenvalues are related to the frequencies~$\left(\omega\right)$ of the obtained normal modes according to $\omega = \sqrt{\lambda}$. The normal modes obtained in this manner are thus QNM's of the liquid. The three smallest eigenvalues are zero as a result of translational invariance of the system, and the remaining $(6N-3)$ eigenvalues have positive definite values.

In Fig. 1(a), we plot the DOS function $G(\omega)$ for the QNM's, by constructing the histogram of the frequencies $\omega$. To improve the statistics we combine the results from all $100$ IS configurations available for a particular $\rho$ and $T$.  To facilitate comparison, the $G(\omega)$ curves for the three densities are normalized so that the area under each is unity.  

The DOS of water has been evaluated previously from simulation data using model potentials other than ST2~\cite{p19,p20,p21a,p21b}.  Our DOS results are consistent with the qualitative pattern found in these earlier studies, both for quenched and instantaneous normal modes.  As in these studies, we find a distinct band of modes below about $400$~cm$^{-1}$ that has been shown to be associated with translational degrees of freedom.  The modes above $400$~cm$^{-1}$ have been assigned to rotational degrees of freedom.  We also note that the qualitative form of $G(\omega)$ varies little across the three densities studied here; this reflects the predominance of the tetrahedral network at all three densities.  At the same time, given the changes in thermodynamic and dynamical properties with density, the nature of the fluctuations around the mean DOS must be separately examined (see next Section) to check for influences connected to density change.

A distinguishing feature of our data for $G(\omega)$ is the gap in the DOS at approximately $850$~cm$^{-1}$.  This feature is absent, or only present as a weak shoulder in the data for other potentials.  In a recent study of SPC/E water, this feature was identified as due to a separation of the rotational modes associated with each of the principal axes of the water molecule, with two axes contributing to the band between $400$ and $850$~cm$^{-1}$, and the third contributing to that above $850$~cm$^{-1}$~\cite{p22}.  Our results suggest that this separation is even more prominent in ST2 water, but further study is required to clarify this point.

Fig.~1(b) shows $p(\omega)$, the average participation ratio as a function of $\omega$, calculated using the eigenvectors for each of the densities. For an individual normal mode $\alpha$, the definition of the participation ratio used here is,
\begin{equation}
p_\alpha=\left[N \sum_{i=1}^{N}\left({\bf{e}}_{\alpha}^{i} \cdot {\bf{e}}_{\alpha}^{i}\right)^{2}\right]^{-1},
\end{equation}
where ${\bf e}_{\alpha}^{i}$ is the vector contribution of the degrees of freedom of particle $i$ to the normalized eigenvector ${\bf e}_{\alpha}$; that is, ${\bf e}_{\alpha}=\sum_{i=1}^N{\bf e}_{\alpha}^{i}$ .
The average participation ratio $p(\omega)$ is the average of $p_\alpha$ for normal modes with eigenfrequencies within $\delta \omega$ of a given $\omega$, where $\delta \omega$ is the bin width used to numerically evaluate $p(\omega)$.  Defined in this way, $p(\omega)$ measures the average fraction of molecules in the system involved in modes having a given frequency.  Extended modes (i.e. modes involving an extensive number of molecules) will have values of the participation ratio that are independent of system size, while localized modes will have values that scale with the system size as $1/N$~\cite{p15}.  

To test which modes are extended or localized, we present in Fig.~1(c) the participation ratios at $\rho = 0.83$~g/cm$^3$ and $T=260$~K obtained from simulations of two different system sizes, $N=216$ and $1728$ molecules.  This data is presented on a semi-log scale in order to reveal the behavior of the data at small values of the participation ratio.  Since these two systems differ in size by a factor of $8$, localized modes in the larger system should have participation ratios approximately $1/8$ of those in the smaller system.  We find that differences on this scale only occur in the gaps between the bands in the DOS occurring at $400$ and $850$~cm$^{-1}$.  Hence, even though the participation ratios near the maximum of the highest frequency band are as low as $0.3$, these still have behavior consistent with extended modes of the system.  This analysis illustrates that the magnitude of the participation ratio alone can be a misleading indicator of whether or not modes at a particular frequency are extended or localized.  

\section{Fluctuations}

\begin{figure}
\centerline{\includegraphics[width=2.6in]{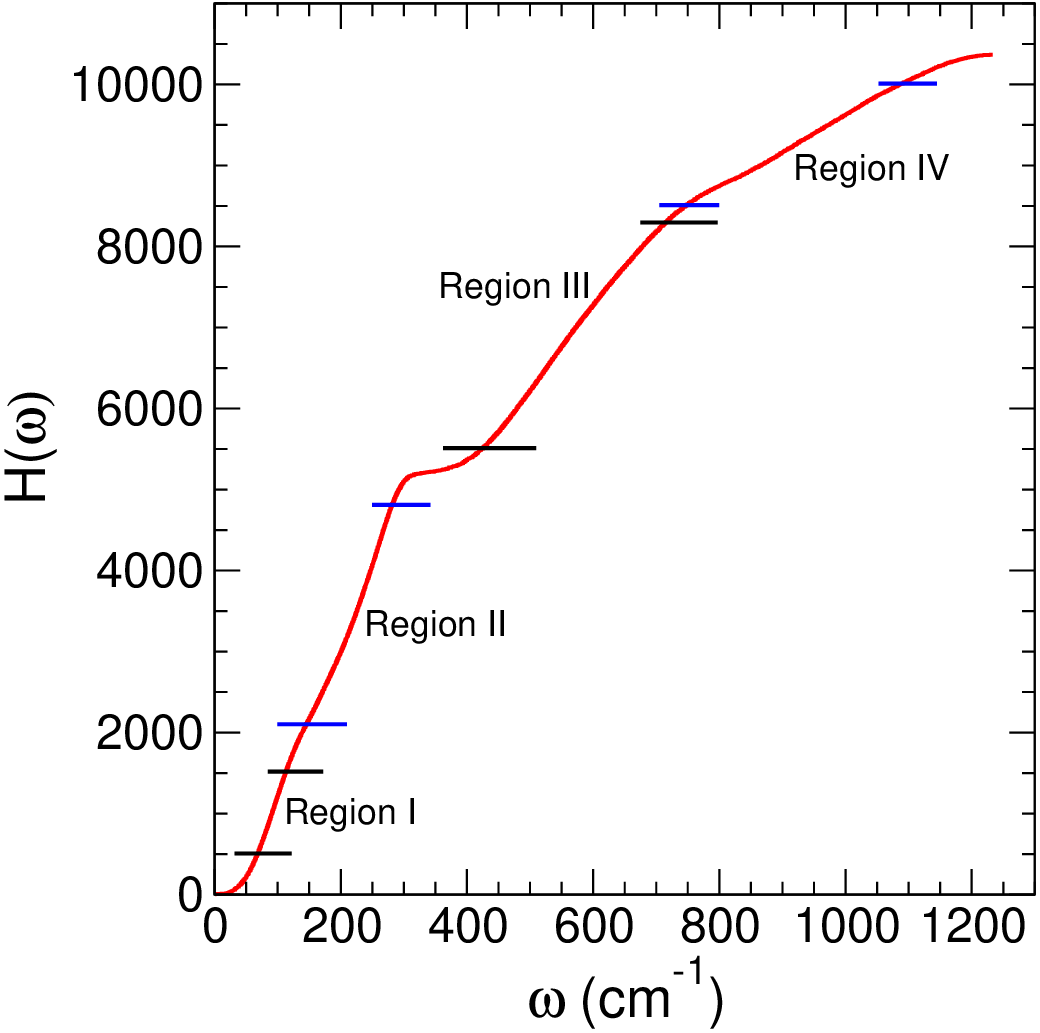}}
\caption{The integral $H(\omega)$ of the density of states for a single spectrum for $\rho = 0.83$~g/cm$^3$ at $T = 260$~K.  Also indicated are the four regions within which a quadratic polynomial is fitted to this function.}
\label{fig2}
\end{figure}

Our next aim is to examine the statistical fluctuations around the average DOS.  Unless stated otherwise, the fluctuation properties presented below are computed for the eigenfrequencies, in contrast to earlier studies~\cite{p10,p11,p12,gsm} that used the eigenvalues.  For a particular IS, we denote the elements of the obtained spectra of eigenfrequencies by ${\omega_\alpha}$ with $\alpha = 1,2,\ldots, 6N$. Since the present system is periodic, the first three elements in the spectrum are zero as they are associated with the translations of the entire system along the three Cartesian directions, and the remaining $(6N -3)$ positive frequencies are characterized by defining a mean local density as well as fluctuations around it. 

In order to study the fluctuation properties, we first must transform the frequencies in such a way that the average spacing between two successive frequencies in a given spectrum is unity. This process is known as ``unfolding" the data~\cite{p10,p11,p12,gsm,p13,p28,p29}.  To unfold the data, we define $H(\omega)$ to be the number of frequencies equal to or less than $\omega$ (shown in Fig. 2), and $S(\omega)$ as a smooth function that passes through $H$ in a best-fit sense.  Note that, as also found in Refs.~\cite{p10,p11,p12}
 in the present case there is no simple function that passes smoothly through the whole of $H(\omega)$.  In order to overcome this complication, we divide the complete spectrum into four regions as shown in Fig. 2. Within each region, we use a quadratic polynomial $D(\omega) = a + b~\omega + c~\omega^2$ as an approximation for $S(\omega)$. The values of $a$, $b$, and $c$ are obtained
by a standard least-squares fitting procedure.  We also calculate the misfit (or residual) function
~\cite{p10,p11,p12} corresponding to the fits in each region to check how well $D(\omega)$ 
approximates $H(\omega)$.  The plots of the misfit functions in our case are qualitatively similar to the plot in Fig. 1(b) of~\cite{p11}.  In order to further improve the fit, in each of the four regions we eliminate subregions where the misfit function has a very irregular behavior. In the remaining regular subregions we fit a new quadratic function to the misfit function and correct $D(\omega)$ by these quadratic functions, which yields the desired unfolding functions. 
For each of the fluctuation properties reported here, we combine the data from all the fitted subregions of all the spectra for a given density and temperature. 

\begin{figure}
\centerline{\includegraphics[width=3.0in]{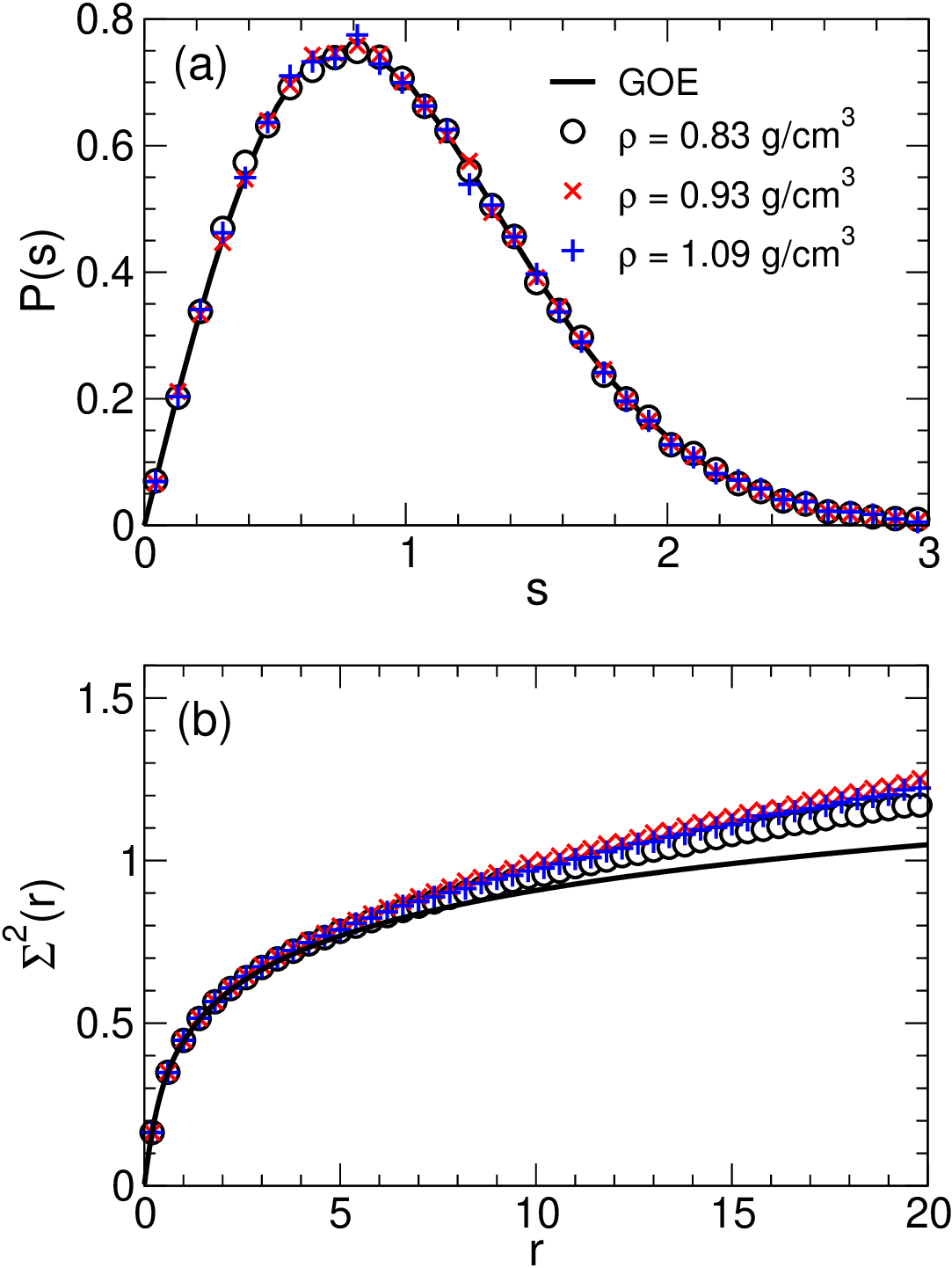}}
\caption{(a). Probability density $P(s)$ of normalized nearest neighbor level spacings $s$ for
several densities at $T = 260$~K. (b). Variance of the number of levels in intervals of length $r$ for several densities at $T = 260$~K.  Symbols are the same as shown in (a). In both (a) and (b), the data are compared to the prediction of the GOE (solid line).}
\label{fig3}
\end{figure}

\begin{figure}
\centerline{\includegraphics[width=3.0in]{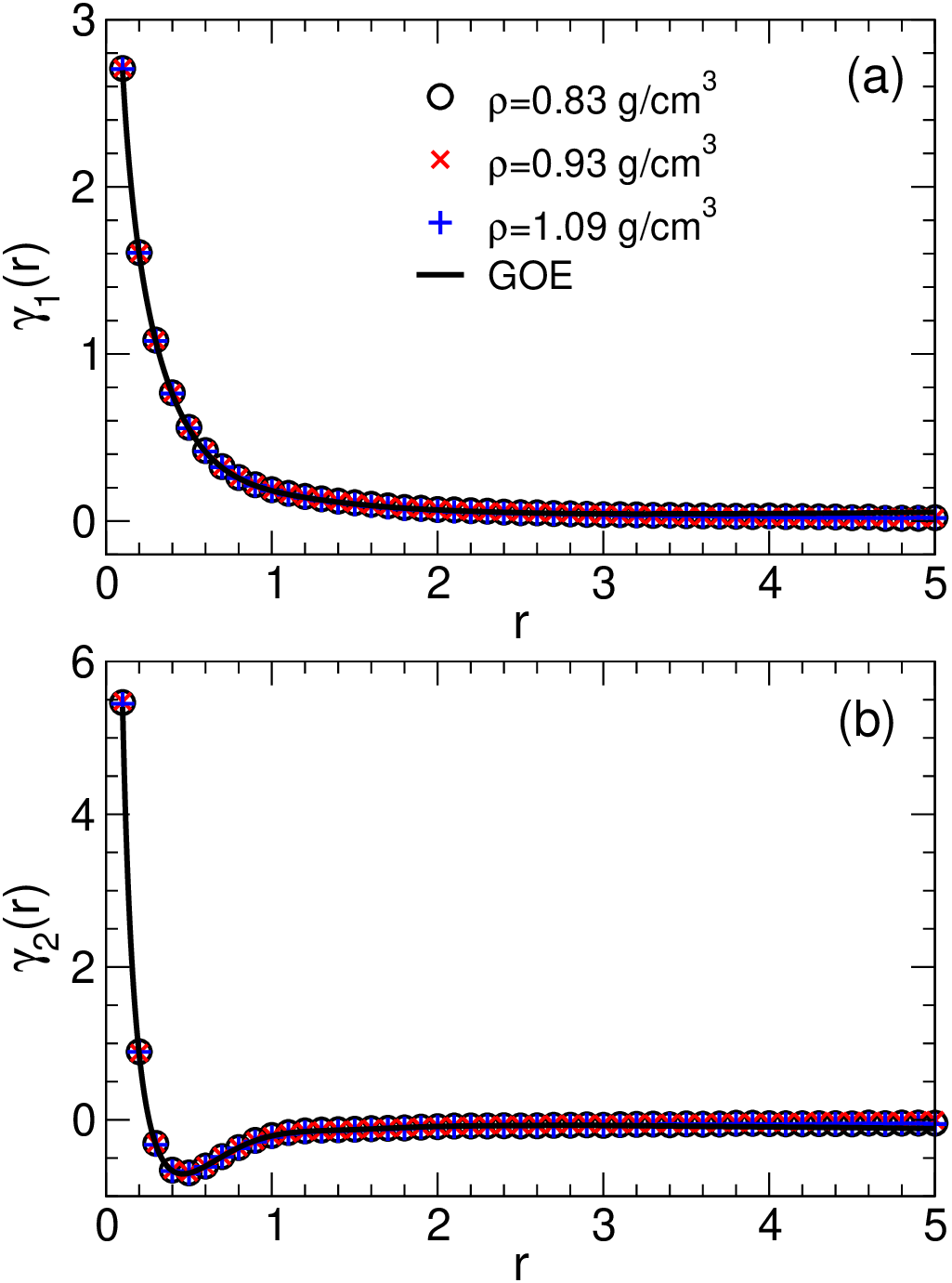}}
\caption{(a) Skewness parameter $\gamma_1(r)$, and (b) excess parameter $\gamma_2(r)$
of the distribution $n\left(r\right)$, the number of levels in intervals of length $r$, shown for several densities at $T = 260$~K.  Symbols in (b) are the same as shown in (a).  In both cases, the data are compared to the prediction of the GOE (solid line).}
\label{fig4}
\end{figure}

The first fluctuation property that we report is the distribution $P(s)$ of the normalized nearest-neighbor spacings $s$ of the frequencies of the unfolded spectra [Fig.~3(a)]. To improve the statistics for each of the three densities studied at $T=260$~K, we combine the spacing data from all four subregions, and from all of the unfolded individual spectra (one for each of the 100 IS configurations).  Also shown in Fig.~3(a) is the prediction based on the GOE of random matrix theory~\cite{p28,p29}.  The agreement of the theory and the data is extremely close.  In addition, we also check that selection of a random individual spectrum and analysis of each of the four regions separately still shows that the fluctuation properties correspond to those of the GOE.

The second fluctuation property we report is $\Sigma^2(r)$, the variance of the number of levels $n(r)$ within an interval of length $r$ located randomly in a given unfolded spectrum. This is plotted for $r$ as large as 20 in Fig. 3(b). It must be emphasized that the quadratic correction applied to the fitting 
function $D(\omega)$ is very important in calculating $\Sigma^2(r)$.  This calculation is extremely sensitive even to very small errors in the aproximation to $S$. The contribution of any such error to $\Sigma^2(r)$ grows as $r^2$, whereas the GOE prediction for $\Sigma^2(r)$ grows only as $\ln(r)$. As described in Ref.~\cite{p11,p12}, even though the broad contour of the misfit fuction is the same for all spectra, the exact locations of the irregular domains vary from one spectrum to another and this might be the cause of the observed deviation observed towards $r = 20$. A detailed analysis using a choice of subdomains tailored to each individual spectrum~\cite{p11,p12} would help to clarify this issue. At the same time, we emphasize that the agreement with theory shown in Fig.~3(b) is of the same quality as observed in the case of amorphous clusters~\cite{p10,p11,p12}.

In Figs. 4(a) and (b), we plot the skewness parameter,
\begin{equation}
\gamma_{1}\left(r\right) 
= \frac{1}{M}\sum^{M}_{j=1}\left[\left(n_{j}-\bar{n}\right)/\sigma\right]^{3},
\end{equation}
and the excess parameter~\cite{p30},
\begin{equation}
\gamma_{2}\left(r\right) = \frac{1}{M}\sum_{j=1}^M\left[\left(n_{j}
-\bar{n}\right)/\sigma\right]^{4} - 3,
\end{equation}
of the $n(r)$ function.  Here $M$ is the number of data values $n_j$ sampled from the $n(r)$ functions for individual unfolded spectra. $\sigma$ and $\bar{n}$ are respectively the standard deviation and mean of the $M$ values of $n(r)$ sampled.  Also included on the same plots are the predictions for the GOE.  These predictions are calculated on the basis of a large ensemble of $500\times500$ matrices belonging to the GOE. We find extremely close agreement with the theoretical prediction for each of the densities.

\begin{figure}
\centerline{\includegraphics[width=3.0in]{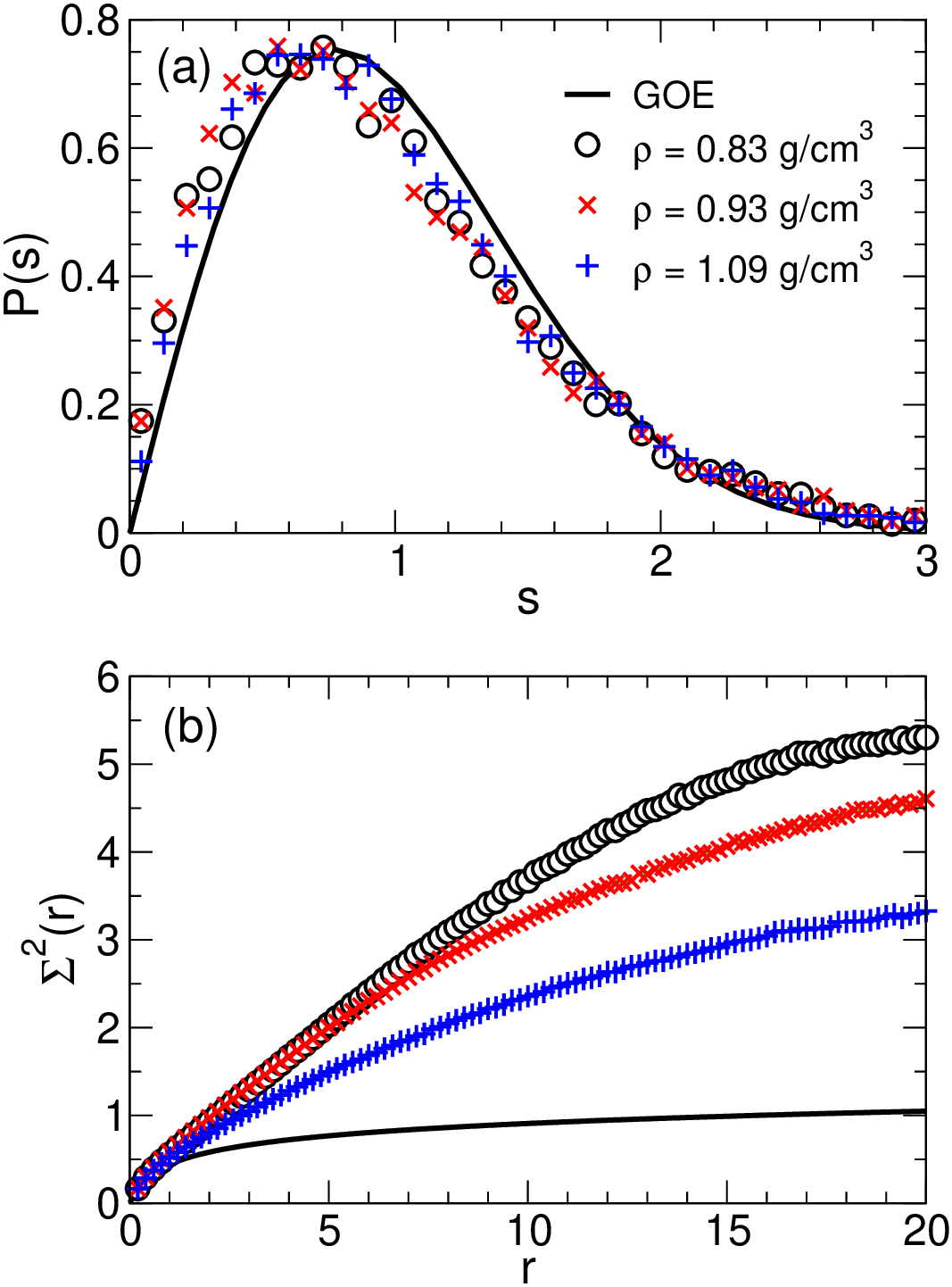}}
\caption{(a). Probability density $P(s)$ of normalized nearest neighbor level spacings $s$ for
several densities at $T = 260$~K, calculated for the subregions in the vicinity of 400 and 800~cm$^{-1}$ (b). Variance of the number of levels in intervals of length $r$ for several densities at $T = 260$~K, calculated for the subregions near 400 and 800~cm$^{-1}$.  Symbols are the same as shown in (a). In both (a) and (b), the data are compared to the prediction of the GOE (solid line).}
\label{fig5}
\end{figure}

Finally, we recall the indication from Fig.~\ref{fig1}(c) that the modes in the vicinity of 400 and 800~cm$^{-1}$ do not involve an extensive number of particles, and are therefore localized modes.  If so, the spectral fluctuations in these subregions should not conform to the GOE predictions.  To test this we have evaluated both $P(s)$ and $\Sigma^2(r)$ using the same procedure as used to create the data in Fig.~\ref{fig3}, except we instead take the spectral data from the region between ``Region II" and ``Region III", and between ``Region III" and ``Region IV", identified in Fig.~\ref{fig2}.  The results are shown in Fig.~\ref{fig5}.  While the data are noisier than in Fig.~\ref{fig3} because of the reduced statistics, there is a systematic departure from the GOE prediction.  While the data do not follow a purely Poissonian form, the peak position of $P(s)$ is shifted in the direction expected when Poissonian effects contribute to the statistics~\cite{new}.  The data are thus consistent with the trend toward localization of the modes in these spectral subregions.

\section{Discussion}

Previous studies on simple liquids, both for clusters and in bulk, and for two and three dimensions, suggest that GOE behavior in the statistics of the DOS is ubiquitous.  Our results show that this commonality also
extends to a molecular liquid across several densities, each representative of a distinct regime of complex liquid behavior.  In all the regimes we examine, neither random tetrahedral network formation, nor proximity to a liquid-liquid critical point, nor entering a higher density regime of fragile glass-forming behavior, is sufficient to induce a departure in the statistics of the DOS from the predictions of the GOE of random matrix theory. Our work therefore considerably extends the evidence in support of the robustness and universality of this behavior in liquid systems. 

We note that the Hessian matrix we obtain is subject to constraints among its elements that have their origin in the conservation of momentum in the associated molecular system.  Such a Hessian matrix is distinct in character from one not subject to a momentum conservation constraint, such as the Hessian associated with an electronic DOS in a disordered system.  The differences in the eigenvalues and eigenvectors in these two cases has been discussed in Refs.~\cite{mezard,stratt1,stratt2}.  In particular, a recent analysis indicates that real-symmetric random matrices subject to momentum conservation may be fundamentally associated with the GOE ensemble~\cite{ma}.  Our results are largely consistent with this analysis.

Our results also shed light on the nature of localization of the normal modes of disordered systems.  While the participation ratios found across a significant fraction of the frequency domain fall below 0.6, analysis of the scaling with system size suggests that the vast majority of the normal modes (excepts those in the ``gaps") are extended in nature, rather than localized.  This is consistent with the fact that these modes obey the statistics of the GOE.  Hence we have evidence that modes with values of $p(\omega)$ as low as 0.3 are still extended in nature.  This result is consistent with an earlier report on the minimum value of the participation ratio for extended modes in a model soft sphere system~\cite{p14}. It would be interesting to examine the nature of extended modes that involve such a low fraction of particles in the system.  We leave this question for future work.

\section{Acknowledgments}
We are grateful to F.~Sciortino for providing us with his code for evaluating the IS configurations and corresponding Hessian matrices for the ST2 potential, and for helpful comments on the manuscript.  GSM would like to thank Subir K. Sarkar for useful comments and discussions on the work. We also thank ACEnet for providing computing resources.  GSM is supported by an ACEnet Research Fellowship and NSERC.  MSGR is funded by AIF and NSERC.  Funding support for PHP is provided by AIF, NSERC and the CRC program.


\begin{thebibliography}{99}
\bibitem{FN1}  We emphasize that we do not mean to imply that GOE behavior is not found in crystal-forming liquids.  The focus on glass-forming liquids is simply a reflection of the significance of this behavior for understanding the glass transition.
\end{thebibliography}
\end{document}